\documentclass[twocolumn,floatfix,showkeys]{revtex4-1}
\usepackage{color}
\usepackage{graphicx}
\usepackage{natbib}
\usepackage{epsfig}
\usepackage{setspace}
\usepackage{amsmath,amssymb}
\usepackage{verbatim}

\newcommand{\diff}{\ensuremath{\mathrm{d}}}
\newcommand{\sech}{\ensuremath{\mathrm{sech}}}

\newcommand{\ga}{\ensuremath{\gamma}}
\newcommand{\ep}{\ensuremath{\epsilon}}
\newcommand{\ebar}{\ensuremath{\bar{\epsilon}}}
\newcommand{\gbar}{\ensuremath{\bar{\gamma}}}

\newcommand{\vbar}{\ensuremath{\bar{V}}}

\begin{document}
\title{Protocols for characterising quantum transport through nano-structures 
       }
\author{Sudeshna Sen}
\affiliation{Chemistry and Physics of Materials Unit, Jawaharlal Nehru 
             Centre for Advanced Scientific Research, Jakkur P.O, 
             Bengaluru 560064, India.}

\author{N.\ S.\ Vidhyadhiraja}
\affiliation{Theoretical Sciences Unit, Jawaharlal Nehru Centre for 
             Advanced Scientific Research, Jakkur P.O, Bengaluru 560064, 
             India.}
\email[]{raja@jncasr.ac.in}
\date{\today}
\begin{abstract}
In this work, we have analysed the exact closed-form solutions for 
transport quantities through a mesoscopic region which 
may be characterised by a polynomial functional of resonant transmission 
functions. These are then utilized to develop considerably improved 
protocols for parameters relevant for quantum transport through molecular 
junctions and quantum dots. The protocols are shown to be experimentally 
feasible and should yield the parameters at much higher resolution
than the previously proposed ones.
\end{abstract}
\maketitle


The single particle scattering approach, pioneered by Landauer 
\cite{Landauer1957,Landauer1970}, 
and extended by B$\mathrm{\ddot{u}}$ttiker \cite{Buttiker1986} 
has been the most extensively employed framework for 
investigating quantum transport through nano-structures. A steady-state 
non-equilibrium problem is mapped to a scattering problem in this approach. 
Realising the importance and applicability of such an approach, a Landauer 
like formula was derived by Meir and Wingreen \cite{MeirWingreen1992} for 
an interacting mesoscopic region coupled to non-interacting leads with 
coupling strengths of ${\mathbf{\Gamma_L}}$ and ${\mathbf{\Gamma_R}}$ to 
the left and right lead respectively. 
For ${\mathbf{\Gamma_L}}=\lambda{\mathbf{\Gamma_R}}$, the current through 
the region flowing into one of the leads may be expressed as an 
integral transform, namely, 
\begin{equation}
  I(V)=\int_{-\infty}^{\infty} K(\ep;V) L(\ep)\diff\epsilon 
  \label{eq:c1}
\end{equation}
in terms of the local properties of the region ($L(\ep)$)
by a kernel $K(\ep;V)$ over the real line. The kernel is given by
$K(\ep;V)=f(\ep-eV/2;T_L) - f(\ep +eV/2;T_R)$, with
$
f(x;T)=(e^{x/k_BT}+1)^{-1}
$
as the Fermi-Dirac distribution function; $T_L$ and $T_R$ are 
the temperatures of the left and right reservoirs respectively,
and V, distributed symmetrically across the two electrodes as 
the voltage bias. The quantity representing the device region is
$L(\ep)$ and for interacting systems, is represented in terms of 
full non-equilibrium Greens functions (retarded $(\mathbf{G^r})$ 
or/and advanced $(\mathbf{G^a})$) 
of the interacting device region, in close resemblance with the 
Landauer formula \cite{MeirWingreen1992}, as 
$
L(\ep)=\frac{-2e}{h}\mathrm{Im\lbrack tr}\lbrace\mathbf{\Gamma G^r
}\rbrace\rbrack,
$
where, $\mathbf{\Gamma}=\mathbf{\Gamma_L \Gamma_R/(\Gamma_L+\Gamma_R)}$.
Even if ${\mathbf{\Gamma_L}}\ne\lambda{\mathbf{\Gamma_R}}$, it was shown by 
Ness \textit{et.~al.}\cite{Ness2010} that the current can be written as 
eq~(\ref{eq:c1}) with renormalized coupling to the electrodes. Such a 
Landauer-like approach is suitable for any mean-field based method, 
for example, density-functional-based techniques (DFT, TDDFT) or 
interactions treated at the Hartree-Fock level \cite{Ness2010}.
For a non-interacting region, eq~(\ref{eq:c1}) reduces to the Landauer 
formula in which 
case the $L(\ep)$ is simply the transmission function of the device under 
consideration.

For a specific form of $L(\ep)$, such as a resonant 
transmission function (RTF) (a Lorentzian),
an exact solution of eq~(\ref{eq:c1}) exists~\cite{Galperin2003,DigammaBulka}.
Thus, if the $L(\ep)$ can be represented as a polynomial 
functional of RTFs, such as a linear superposition of 
multiple (Lorentzian) resonances (at the lowest order)
\begin{equation}
  L(\epsilon)=\sum_r\frac{A_r}{(\epsilon-\epsilon_r)^2+\gamma_r^2}\,,
\label{eq:c2}
\end{equation}
the exact solution of eq~(\ref{eq:c1}) can be obtained 
(with the use of partial fractions for higher orders). 
Such a form has indeed been found for a variety of nano-systems 
such as molecular junctions, quantum dots and quantum point contacts 
\cite{Johnson1992,Foxman1993,Reddy2007,CNTSET,Linke2012}, where the 
discrete level $\ep_r$ gets broadened due to the coupling ($\gamma_r$) 
to the macroscopic leads, and $A_r$ is an unknown parameter that 
depends on a number of factors like the number of conducting molecules 
in the junction or the coupling with the electrodes \cite{Chen2010}. 
More importantly, it is not restricted to systems in the ballistic 
regime and is applicable even for interacting mesoscopic systems
under a broad range of experimentally relevant conditions
\cite{Stafford1996,Johnson1992,Foxman1993,Linke2012}. 
As mentioned in \cite{Stafford1996}, such a Breit-Wigner (BW) type 
resonant conductance formula is relevant for an interacting system 
with a non-degenerate ground state like in semiconductor nanostructures 
or ultra small metallic systems. The positions and intrinsic widths 
of the BW type resonances are determined by the many body states 
of the interacting system. The magnitudes of the temperature 
$(k_BT)$, bias $eV$ and coupling to the electrodes should be much 
smaller than the resonant energy so that only a single transition 
from the $N$ electron ground state (GS) to 
$N+1$ electron GS is allowed. At a finite voltage these 
resonances may get shifted \cite{Christen1994,Stafford1996,Chen2010} 
relative to the zero bias position. This shift manifests itself 
differently for a symmetric or asymmetric junction. With minor 
change of $\ep_r \rightarrow \ep_r+\eta V$, asymmetric couplings to 
the electrodes may also be incorporated with $\eta=0$ being the 
symmetric case. Naturally, strongly interacting 
systems such as those where Kondo physics is important
exhibit slow logarithmically decaying tails in $L(\ep)$, and 
hence cannot be captured
within such an approach \cite{Logan2001}, since Lorentzians 
have a algebraically decaying tail
structure. 

In this work, we have analysed the exact solutions of eq~(\ref{eq:c1}) with
an $L(\ep)$ given by eq~(\ref{eq:c2}) and discuss their general validity. 
Further, we utilize them in designing substantially improved protocols for 
finding the parameters of the $L(\ep)$, specifically for quantum 
dots and molecular junctions. These protocols are shown to
 be experimentally feasible and if implemented, will yield the 
parameters with much higher resolution than the existing ones.
An understanding of the non-linear regime, both in terms of voltage 
and thermal bias is important. Such an insight is most easily 
developed through closed form analytical expressions. The asymptotic 
properties of exact solutions are the most straightforward route to 
obtaining such expressions, thus emphasising the utility of exact 
solutions.

Substituting eq~(\ref{eq:c2}) with $A_r=\gamma_r^2$ in eq~(\ref{eq:c1}), 
and transforming it to a contour integration \cite{Galperin2003},  
we get the following expression \cite{DigammaBulka}
for the current $I(V,T_L,T_R)$.
\begin{equation}
  I=\frac{2e}{h}\sum_r\gamma_r\,{\rm Im}\left[\Psi(z_{Lr})-\Psi(z_{Rr})\right]
\label{eq:c3}
\end{equation}
where $\Psi(z)$ is the digamma function \cite{Stegun} and
$z_{L/R,r}=1/2+ [\gamma_r - i (\ep_r \mp eV/2)]/(2\pi k_BT_{L/R})$.
Using the current expression thus obtained, we 
can simply write down the differential conductance 
$G(V,T_L,T_R)=dI/dV$ and differential thermopower 
$S(V,T_L,T_R)=\partial\Delta V / \partial\Delta T$. 
These are given by 
\begin{equation}
   G=\frac{e^2}{h}\sum_r\gamma_r{\rm Re}\left[\frac{\Psi^\prime(z_{Lr})}{k_BT_L}
                +\frac{\Psi^\prime(z_{Rr})}{k_BT_R}\right]
\label{eq:c4}
\end{equation}
where $\Psi^\prime(z)$ is the trigamma function \cite{Stegun} and
\begin{equation}
  S=-\frac{k_B}{e}\frac{\sum_r\gamma_r{\rm Im}\left[
                  \frac{\lambda_{Lr}\Psi^\prime(z_{Lr})}{k_BT_L}+
               \frac{\lambda_{Rr}\Psi^\prime(z_{Rr})}{k_BT_R}\right]}
               {\sum_r\gamma_r{\rm Re}\left[\frac{\Psi^\prime(z_{Lr})}{k_BT_L}+
               \frac{\Psi^\prime(z_{Rr})}{k_BT_R}\right]}.
  \label{eq:c5}
\end{equation}
An expression for thermopower in terms of trigamma functions has been 
obtained earlier {\em in the linear response regime} \cite{Subroto2008,Rejec2012}. 
The expression derived here is exact and hence represents a generalization 
of that result to all regimes. 

It is worth considering the general structure of the equations above. 
The thermal energy, $k_BT_{L/R}$ sets the reference scale, since the final 
expression is a function only of 
$\ebar_{L/R,r}=\epsilon_{L/R,r}/(2\pi k_BT_{L/R})$,
$\gbar_{L/R,r}=\gamma_r/(2\pi k_BT_{L/R})$ and 
$\vbar=eV/(2\pi k_BT_{L/R})$. The width 
of the resonance appears in the real part of the argument, while
the peak energy and the bias voltage appear in the imaginary part 
of $z_{L/R}$. Since many of the properties of the digamma function 
\cite{identities_SI} depend on the real and 
imaginary parts separately, we can classify the parameter space 
into the following regions: narrow resonance ($\gbar_r\ll 1$), 
finite width resonance ($\gbar_r\sim 1$) and broad resonance 
($\gbar_r\gg 1$). 

\paragraph{Narrow resonance:}
The narrow resonance regime is characterized by having the width of the 
resonance being much smaller than the thermal energy scale 
($\gbar_r\ll 1$). 
In the expression given by equation~(\ref{eq:c3}),
if $\bar{\gamma}_r\ll 1$, then using the identity, 
${\rm Im}\Psi(1/2+iy)=(\pi\tanh \pi y)/2$ \cite{Stegun}, we 
get
\begin{equation}
  I=\frac{e\pi}{h}\sum_r\gamma_r\left[
     \tanh\left(\frac{\alpha_{Rr}}{2}\right)-
     \tanh\left(\frac{\alpha_{Lr}}{2}\right)\right].
\label{eq:I_NR}
\end{equation}
where  $\alpha_{L/R,r}=(\epsilon_r \mp eV/2)/k_BT_{L/R}$.
A similar result has been obtained before, through Keldysh 
formalism \cite{KopninGalperin2009} for resonant transmission through one 
dimensional systems. 
The differential conductance 
in this regime can be obtained by calculating $G=dI/dV$ and is 
given by,
\begin{equation}
  G=\frac{e^2\pi}{4k_Bh}\sum_r\gamma_r\left[
    \frac{\sech^2\left(\frac{\alpha_{Lr}}{2}\right)}{T_L}+
    \frac{\sech^2\left(\frac{\alpha_{Rr}}{2}\right)}{T_R}\right],
\label{eq:narrow_G}
\end{equation}
which yields conductance oscillations as a function of source-drain bias 
(when $eV=\pm 2\ep_r$) or as a function of gate voltage (which tunes $\ep_r$) 
at zero bias \cite{Beenakker1991,CNTSET}.
It is well known that the positions of resonances in the zero bias 
conductance versus gate voltage curve yields values of the resonance 
energies. The interpretations of these oscillations as being due to 
resonant transmission or due to Coulomb blockade rests on the dependence 
of the energy level spacing $\Delta E_r=\ep_{r+1}-\ep_r$ on the bias. If 
the spacing increases with increasing bias, then the energies are 
single-particle energies, while for constant spacing, the levels are 
many-particle levels that include charging energy 
\cite{Stafford1996,Linke2012}.

If the individual resonance peaks are separated by energies far greater 
than the thermal energy scales ($k_BT_{L/R}$) and the widths ($\gamma_r$), 
then the resonance closest to the chemical potential would be the major 
contributor to the current, and hence a single resonance TF may be assumed
with width $\gamma$ and peaked at $\ep_0$. 
Such a situation may be realized in quantum dots by reaching sufficiently 
low temperature. In such a case, we can obtain the well known expression 
for thermovoltage, $V_{th}$, as,
$
  V_{th}=-S_{ideal}\Theta,
$
where, $S_{ideal}= \ep_0/(eT)$ is the thermopower of an ideal quantum dot 
characterised by a zero width ($\delta$- function) TF, $T$ is the 
average temperature given by $T=(T_R+T_L)/2$ and $\Theta=T_L-T_R$ is the 
thermal bias. 

In a recent work by Mani \textit{et.~al.} \cite{Linke2011}, 
a protocol for obtaining the width, $\gamma$, of an RTF was 
proposed through the measurement of a quantity termed thermopower 
offset ($F$) defined as 
$
  F=(S_{ideal}-S)/S_{ideal}
$.  This quantity 
was shown, employing numerical calculations, to be a simple polynomial 
function of $\gamma/k_BT$ (when $\ep_0\rightarrow 0$), i.e.\
$F=A(\gamma/k_BT)+B(\gamma/k_BT)^2$ with $A$ and $B$ being constants.
By an experimental measurement of $F$, the above 
equation can be inverted to find $\gamma/k_BT$.

Since this protocol relies on an accurate measurement of $S$ in the 
limit $\ep_0\rightarrow 0$, where $S$ would itself be vanishingly small, 
$F$ would be a difficult quantity to measure with high resolution.  
Hence we propose an alternative protocol for finding $\gamma$, which 
does not require tuning of the resonant level to zero. We first state 
that the thermopower of a quantum dot in the linear response regime and 
in the limit $\ep_0\rightarrow 0$ may be represented as, 
\begin{equation}
S_{protocol}=S_{ideal}\frac{A+B\gamma/k_BT}{A+C\gamma/k_BT}
\label{eq:Sprotocol}
\end{equation}
where $A,B$ and $C$ are pure constants 
\cite{NR_SI}, given by
$A=3\zeta(2)$, $C=14\zeta(3)$ and $B=2C$ 
($\zeta(z)$ is the Reimann zeta function), which implies that the
 linear term in the offset expression has the coefficient
$7\zeta(3)/3\pi\zeta(2)\simeq 0.54$, which matches with the value 
obtained by Mani et al \cite{Linke2011} through numerical fitting,
and also shows that these coefficients are indeed pure numbers. 
For $\ep_0\neq 0$, we obtain a general expression for 
the thermopower \cite{NR_SI}, given by,
\begin{equation}
  S_{protocol}=\frac{k_B}{e}\frac{\bar{F}_1}{\bar{F}_0}
               \left(\frac{1+\frac{\bar{C}_1}{\bar{F}_1}\frac{\gamma}{k_B}}
               {1+\frac{\bar{C}_0}{\bar{F}_0}\frac{\ga}{k_B}}\right).
  \label{eq:SNL}
\end{equation} 
The quantity $\frac{\bar{F}_1}{\bar{F}_0}$ 
is the general ideal thermopower (in units of $k_B/e$), that reduces 
to $\epsilon_0/eT$ in the linear response regime. The quantities 
$\bar{C}_0$ and $\bar{C}_1$ are functions purely of $\epsilon_0$, 
$T_L$ and $T_R$ and can be expressed in terms of polygamma functions. 
These may be easily evaluated either using the series expansions given 
in the SM \cite{NR_SI} or through technical software like 
MATLAB \cite{MATLAB}. 
Thus the protocol simply 
consists of measuring thermopower at a given $\ep_0$, which can be chosen 
such that high resolution is achieved; finding the coefficients 
$\bar{F}_1$, $\bar{F}_0$, $\bar{C}_0$ and $\bar{C}_1$ using the 
expressions given in SM \cite{NR_SI}, and inverting equation~(\ref{eq:Sprotocol}) or 
equation~(\ref{eq:SNL}) to get the resonance width $\gamma$.
\begin{figure}
\includegraphics[clip=,scale=0.35]{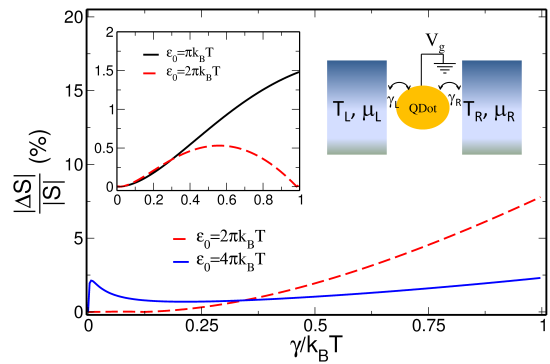}%
  \caption{Main panel: The relative discrepancy in thermopower, 
           $\Delta S/S$, defined as $(S_{exact}-S_{protocol})/S_{exact}$
(computed using equations~\ref{eq:c5} and ~\ref{eq:SNL}) 
           is plotted as a function of $\gamma/k_BT$ 
           for non-zero $\epsilon_0=4\pi k_BT$ (blue solid line) 
           and $2\pi k_BT$ (red dashed line) at $T=3$ K and $\Theta=0.1$ K. 
           Inset: The same quantity is calculated in the nonlinear 
           regime at $T=3$ K and $\Theta=3$ K \cite{NL_parameters}. 
           The relative discrepancy 
           reduces to 1-2\% in the nonlinear regime thus offering an 
           improved standard. The schematic represents the relevance of this
	   protocol for quantum dot  devices.}
  \label{fig:S}
\end{figure}

We have bench marked the expression used for this protocol in
 Fig.~(\ref{fig:S}), where we show the difference
between the protocol expression equation~(\ref{eq:SNL}) and the 
exact result equation~(\ref{eq:c5}) as a function of the scaled 
width $\gamma/k_BT$ for various resonant level positions. It is seen 
that the discrepancy, $\Delta S/S$, between the exact (equation~(\ref{eq:c5})) 
and the protocol expression (equation~(\ref{eq:SNL})) is less than 4\% 
for an $\ep_0$ as large as $2\pi k_BT$. In fact an analysis of 
thermopower tells us that the discrepancy, $\Delta S$ 
reduces dramatically in the nonlinear regime thus offering an even 
better protocol. This is seen in the inset of fig.~(\ref{fig:S}) where 
$\Delta S$ is seen to decrease in the presence of a finite thermal 
bias (as compared to the main panel). We have seen that the discrepancy 
between the protocol and exact thermopower offset near 
$\ep_0\rightarrow 0$ can be as large as $20\%$, and is hence less 
reliable (see Figure~(4) in SM \cite{NR_SI}). 
\paragraph{Broad Resonance:} 
The broad resonance transmission defined by $\gamma\gg k_BT_L,$ $k_BT_R$  
is most appropriate for molecular junctions where the HOMO and LUMO levels 
are broadened due to the coupling with the leads, and the width of these 
levels could easily be of the order of eV, which is far greater than the 
thermal energy scales. Further, since the HOMO-LUMO level separation is 
much greater compared to the width or the thermal energy scale, a single 
resonance $L(\ep)$ can again be assumed. It is easy to see that in 
equation~(\ref{eq:c3}), 
if $\gbar_{L/R}\gg 1$, then $|z_{L/R}|\gg 1$ irrespective of the values 
of the $\ep_0$ or the bias, since the latter are present in the 
imaginary part, while $\gamma$ is in the real part of $z_{L/R}$.
This allows us to use the asymptotic form of digamma function \cite{Stegun} 
for large $z$, which is
$\Psi(z)\sim \ln z - 1/z; z\rightarrow \infty, |{\rm arg}(z)|<\pi$.  
Hence, we get for $I$ in units of $2e/h$, 
\begin{align}
  I=&\sum_r\bigg\lbrace\gamma_r\left[\mathrm{tan}^{-1}
     \left(\frac{-\lambda_{r-}}{\gamma_{r}}\right)
    +\mathrm{tan}^{-1}\left(\frac{\lambda_{r+}}{\gamma_{r}}\right)
     \right]\nonumber\\
    +&\frac{\gamma_r^2\pi^2k_B^2}{3}\bigg[\frac{T_L^2\lambda_{r-}}
           {(\gamma_r^2+\lambda_{r-}^2)^2}
    -\frac{T_R^2\lambda_{r+}}{(\gamma^2+\lambda_{r+}^2)^2}\bigg]
     \bigg\rbrace,
  \label{eq:I_BR}
\end{align}
where,
$\lambda_{r,\pm}=\epsilon_r\pm eV/2$.
As the above arguments show, this equation is valid for arbitrary values 
of the resonance position $\ebar$ or the voltage bias $\vbar$ as long as 
$\gbar\gg 1$ is satisfied implying that it is sufficient for the thermal 
energy scales to be much smaller compared to the resonance width. Although 
the above arguments seems to be based upon a broad resonance condition, a 
subtle point to note is that this form may be applied for arbitrary widths 
($\gamma/k_BT$) if the magnitude of the level position (measured from the 
chemical potential) is large compared to $\mathrm{max}(eV/2,k_BT)$, 
thus making it useful for molecular junctions. The result obtained here 
represents a generalization of an expression obtained by 
Stafford \cite{Stafford1996,Ioan2012}, (for RTFs in the wide 
band approximation) at $T=0$ to finite temperatures and a 
finite thermal bias. We can now clearly see through eq~(\ref{eq:I_BR}), 
the emergence of a temperature controlled current rectifier. 
This rectification current, defined as 
$\Delta I= (I(V,T_L,T_R)+I(-V,T_L,T_R))/2$ in units of $\frac{2e}{h}$ is 
given by,
\begin{equation}
  \Delta I=\sum_r\frac{\gamma_r^2\pi^2k_B^2}{3}T\Theta\left[
           \frac{\lambda_+}{(\gamma_r^2+\lambda_+^2)^2}+
           \frac{\lambda_-}{(\gamma_r^2+\lambda_-^2)^2}\right]
  \label{eq:DI}
\end{equation}
This rectification current is experimentally measurable ($\sim$~nA) 
(Figure~(3) in SM \cite{BR_SI}) 
even at a temperature and thermal bias of $50$ K each. 
This motivates us to design a protocol for predicting the resonant 
transmission function parameters through $\Delta I$ at zero voltage 
bias (which is in fact the thermocurrent, $\Delta I_{th}$) in conjunction 
with the zero bias conductance. This protocol 
involving the thermocurrent ($\Delta I_{th}$) shall be discussed below.

 The existing protocol for finding the resonant level in molecular junctions is implemented through 
transition voltage spectroscopy (TVS) 
\cite{TVS1Yu,TVS2Liu,Beebe2006,Beebe2008,Tan2010} . 
The basis for this protocol is the 
existence of a minimum, in the $\ln~(I/V^2)$ versus $1/V$ curve.
This minimum is useful because, it occurs at a voltage that is much smaller
and accessible than the resonance condition voltage ($V=2\ep_0$). 
The interpretation of this voltage minimum  as $V_m^{conv}=2\ep_0/\sqrt{3}$ 
relies on a result ~\cite{Ioan2012} obtained in the limit when $\gamma/\epsilon_0\ll1$, and 
hence the information on the width of the resonance 
is completely lost. We have obtained a result for the TVS minimum that is 
valid for a regime where $\gamma/\ep_0$ may be significant 
(e.~g.~amine linked junctions) \cite{Hybertsen2008,Hybertsen2007}, 
which is, 
\begin{equation}
  \frac{eV_m}{\epsilon_0}=\frac{eV_m^{conv}}{\epsilon_0}\left[1+5\left(\frac{\gamma}{\epsilon_0}\right)^2\right]^{1/2}
\,.
  \label{TVS2}
\end{equation}
\begin{figure}
\centerline{\includegraphics[clip=,scale=0.25]{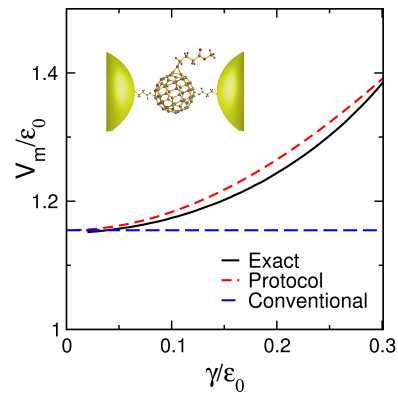}}
  \caption{Comparison of TVS minimum, $V_m$, obtained using 
           the protocol equation~(\ref{TVS2}) (red dashed line), 
           exact solution~\cite{BR_SI}(black solid line) and the 
           conventional~\cite{Ioan2012} $V_m^{conv}=2\epsilon_0/\sqrt{3}$
           (blue dashed line). The schematic in the inset represents a molecular junction. 
\label{fig:TVS}} 
\end{figure}
A comparison between the $V_m$ obtained and 
the $V_m^{conv}$ with the exact result \cite{BR_SI} 
as obtained by numerically finding the minimum is shown 
in fig.~(\ref{fig:TVS}). 
It is seen that the deviation from the conventional TVS minimum occurs even 
at very small widths ($\gamma/\ep_0\sim 0.05$), while the agreement with the exact
result is excellent. Thus the experimentally 
measured $V_m$ would contain information about both $\ep_0$ and $\gamma$
and hence by itself, cannot be used to find the level position and the width.
A second equation is needed that relates an experimentally measurable quantity
to   $\ep_0$ and $\gamma$. We state that such a quantity is the ratio $R_{TVS}$ of
 $\Delta I_{th}$ to the zero bias conductance $G$. The latter 
(obtained either by using 
equation~(\ref{eq:c4}) or using equation~(\ref{eq:I_BR}), 
 for a single resonance with $T_L=T_R=T$ is given by
\begin{equation}
  G=\frac{2e^2\gamma^2}{h}\left[\frac{1}{2(\gamma^2+\lambda_+^2)}+
    \frac{1}{2(\gamma^2+\lambda_-^2)}\right].
  \label{eq:GBR}
\end{equation}
and hence the ratio is given by:
\begin{equation}
  R_{TVS}=\frac{\Delta I_{th}}{G}=\frac{k_B^2\pi^2}{3e}\frac{4T\Theta\epsilon_0}
                          {(\gamma^2+\epsilon_0^2)}\,.
  \label{dI_G}
\end{equation}
Equations~(\ref{TVS2}) and (\ref{dI_G}) can be easily combined to get 
both $\gamma$ and $\epsilon_0$. So if we define $X=\sqrt{3}eV_m/2$ 
and $Y=3eR_{TVS}/(4\pi^2 T\Theta)$, which are experimentally
measurable, then $\gamma$ and $\ep_0$ may be obtained through simple 
expressions involving $X$ and $Y$ \cite{BR_SI}.

\paragraph{Finite width:}
In the final part, we provide expressions for the $I-V$ characteristic 
 in the finite width regime ($\gamma \sim k_BT$), which has hitherto been analytically
inaccessible. 
We can utilize the recurrence relations and the 
multiplication formula for the digamma function \cite{Stegun} to get 
the following result. 
If  $T_L=T_R=T$ and $\gamma_r=n\pi k_BT;\;n=1,2,...$, then the current $I$ 
in units of $2e/h$ is given by,
for $n=2m+1$
\begin{align*}
  I=&\sum_r\bigg[\frac{\pi}{\alpha_{L}}-\frac{\pi}{\alpha_{R}}+
    \frac{\pi}{2}\left(\coth(\pi\alpha_{Lr})-\coth(\pi\alpha_{Rr})\right)+ 
    \nonumber\\
    &\sum_{rk=0}^m\left\lbrace\frac{\alpha_{Lr}}{k^2+|\nu_L|^2}-
    \frac{\alpha_{Rr}}{k^2+|\nu_R|^2}\right\rbrace\bigg],
\end{align*}
where, $\nu_{L/R}=-\frac{i\alpha_{L/R}}{2\pi}$. 
The case for $n=2m$ can also be similarly derived \cite{FW}. 

Finding the $L(\ep)$ for systems where electron-electron and electron-phonon
interactions play a dominant role, is of course a major bottleneck
and is currently a major topic of research 
\cite{Ness2010, Sanvito2012, benzene, Kotliar2010, Markussennanolett, 
Stafford2009, DiVentra}.
Nevertheless, it is clear from our work that expressing the local properties 
of the interacting region ($L(\ep)$) in terms of RTFs allows the utilization 
of exact results. Subsequently,  we have proposed substantially improved 
protocols that can be implemented experimentally for finding transmission 
function parameters in quantum dots and molecular junctions. Our results 
being based on exact solutions also provide analytical insight into
the difficult nonlinear regime, and provide a unified platform for the analysis
of simulations and experiments in quantum transport through nanostructures.

\begin{acknowledgments}
The authors thank Prof. Timothy S. Fisher for fruitful discussions.
The authors acknowledge funding and support from CSIR~(India) and DST~(India). 
\end{acknowledgments}

\bibliography{paper}

\begin{thebibliography}{10}%
\makeatletter
\providecommand \@ifxundefined [1]{%
 \ifx #1\undefined \expandafter \@firstoftwo
 \else \expandafter \@secondoftwo
\fi
}%
\providecommand \@ifnum [1]{%
 \ifnum #1\expandafter \@firstoftwo
 \else \expandafter \@secondoftwo
\fi
}%
\providecommand \enquote [1]{``#1''}%
\providecommand \bibnamefont  [1]{#1}%
\providecommand \bibfnamefont [1]{#1}%
\providecommand \citenamefont [1]{#1}%
\providecommand\href[0]{\@sanitize\@href}%
\providecommand\@href[1]{\endgroup\@@startlink{#1}\endgroup\@@href}%
\providecommand\@@href[1]{#1\@@endlink}%
\providecommand \@sanitize [0]{\begingroup\catcode`\&12\catcode`\#12\relax}%
\@ifxundefined \pdfoutput {\@firstoftwo}{%
 \@ifnum{\z@=\pdfoutput}{\@firstoftwo}{\@secondoftwo}%
}{%
 \providecommand\@@startlink[1]{\leavevmode\special{html:<a href="#1">}}%
 \providecommand\@@endlink[0]{\special{html:</a>}}%
}{%
 \providecommand\@@startlink[1]{%
  \leavevmode
  \pdfstartlink
   attr{/Border[0 0 1 ]/H/I/C[0 1 1]}%
   user{/Subtype/Link/A<</Type/Action/S/URI/URI(#1)>>}%
  \relax
 }%
 \providecommand\@@endlink[0]{\pdfendlink}%
}%
\providecommand \url  [0]{\begingroup\@sanitize \@url }%
\providecommand \@url [1]{\endgroup\@href {#1}{\urlprefix}}%
\providecommand \urlprefix [0]{URL }%
\providecommand \Eprint[0]{\href }%
\@ifxundefined \urlstyle {%
  \providecommand \doi [1]{doi:\discretionary{}{}{}#1}%
}{%
  \providecommand \doi [0]{doi:\discretionary{}{}{}\begingroup
  \urlstyle{rm}\Url }%
}%
\providecommand \doibase [0]{http://dx.doi.org/}%
\providecommand \Doi[1]{\href{\doibase#1}}%
\providecommand \bibAnnote [3]{%
  \BibitemShut{#1}%
  \begin{quotation}\noindent
    \textsc{Key:}\ #2\\\textsc{Annotation:}\ #3%
  \end{quotation}%
}%
\providecommand \bibAnnoteFile [2]{%
  \IfFileExists{#2}{\bibAnnote {#1} {#2} {\input{#2}}}{}%
}%
\providecommand \typeout [0]{\immediate \write \m@ne }%
\providecommand \selectlanguage [0]{\@gobble}%
\providecommand \bibinfo [0]{\@secondoftwo}%
\providecommand \bibfield [0]{\@secondoftwo}%
\providecommand \translation [1]{[#1]}%
\providecommand \BibitemOpen[0]{}%
\providecommand \bibitemStop [0]{}%
\providecommand \bibitemNoStop [0]{.\EOS\space}%
\providecommand \EOS [0]{\spacefactor3000\relax}%
\providecommand \BibitemShut [1]{\csname bibitem#1\endcsname}%
\bibitem{Landauer1957}%
  \BibitemOpen
  \bibfield{author}{%
  \bibinfo {author} {\bibfnamefont{R.}~\bibnamefont{Landauer}},\ }%
  \bibfield{journal}{%
  \bibinfo {journal} {{I}BM.~{J}.~{R}es.~{D}ev}\ }%
  \textbf{\bibinfo {volume} {1}},\ \bibinfo {pages} {233} (\bibinfo {year}
  {1957})%
  \bibAnnoteFile{NoStop}{Landauer1957}%
\bibitem{Landauer1970}%
  \BibitemOpen
  \bibfield{author}{%
  \bibinfo {author} {\bibfnamefont{R.}~\bibnamefont{Landauer}},\ }%
  \bibfield{journal}{%
  \bibinfo {journal} {{P}hilos.~{M}ag.}\ }%
  \textbf{\bibinfo {volume} {21}},\ \bibinfo {pages} {863} (\bibinfo {year}
  {1970})%
  \bibAnnoteFile{NoStop}{Landauer1970}%
\bibitem{Buttiker1986}%
  \BibitemOpen
  \bibfield{author}{%
  \bibinfo {author}
  {\bibfnamefont{M.}~\bibnamefont{B$\ddot{\mathrm{u}}$ttiker}},\ }%
  \bibfield{journal}{%
  \bibinfo {journal} {{P}hys.~{R}ev.~{L}ett}\ }%
  \textbf{\bibinfo {volume} {57}},\ \bibinfo {pages} {1761} (\bibinfo {year}
  {1986})%
  \bibAnnoteFile{NoStop}{Buttiker1986}%
\bibitem{MeirWingreen1992}%
  \BibitemOpen
  \bibfield{author}{%
  \bibinfo {author} {\bibfnamefont{Y.}~\bibnamefont{Meir}}\ and\ \bibinfo
  {author} {\bibfnamefont{N.~S.}\ \bibnamefont{Wingreen}},\ }%
  \bibfield{journal}{%
  \Doi{10.1103/PhysRevLett.68.2512}{\bibinfo {journal}
  {{P}hys.~{R}ev.~{L}ett.}}\ }%
  \textbf{\bibinfo {volume} {68}},\ \bibinfo {pages} {2512} (\bibinfo {year}
  {1992})%
  \bibAnnoteFile{NoStop}{MeirWingreen1992}%
\bibitem{Ness2010}%
  \BibitemOpen
  \bibfield{author}{%
  \bibinfo {author} {\bibfnamefont{H.}~\bibnamefont{Ness}}, \bibinfo {author}
  {\bibfnamefont{L.~K.}\ \bibnamefont{Dash}},\ and\ \bibinfo {author}
  {\bibfnamefont{R.~W.}\ \bibnamefont{Godby}},\ }%
  \bibfield{journal}{%
  \Doi{10.1103/PhysRevB.82.085426}{\bibinfo {journal} {{P}hys.~{R}ev.~{B}}}\ }%
  \textbf{\bibinfo {volume} {82}},\ \bibinfo {pages} {085426} (\bibinfo {year}
  {2010})%
  \bibAnnoteFile{NoStop}{Ness2010}%
\bibitem{Galperin2003}%
  \BibitemOpen
  \bibfield{author}{%
  \bibinfo {author} {\bibfnamefont{M.}~\bibnamefont{Galperin}}, \bibinfo
  {author} {\bibfnamefont{A.}~\bibnamefont{Nitzan}}, \bibinfo {author}
  {\bibfnamefont{S.}~\bibnamefont{Sek}},\ and\ \bibinfo {author}
  {\bibfnamefont{M.}~\bibnamefont{Majda}},\ }%
  \bibfield{journal}{%
  \bibinfo {journal} {{J}ournal of Electroanalytical Chemistry}\ }%
  \textbf{\bibinfo {volume} {550}},\ \bibinfo {pages} {337} (\bibinfo {year}
  {2003})%
  \bibAnnoteFile{NoStop}{Galperin2003}%
\bibitem{DigammaBulka}%
  \BibitemOpen
  \bibfield{author}{%
  \bibinfo {author} {\bibfnamefont{B.~R.}\ \bibnamefont{Bułka}}\ and\ \bibinfo
  {author} {\bibfnamefont{T.}~\bibnamefont{Kostyrko}},\ }%
  \bibfield{journal}{%
  \Doi{10.1103/PhysRevB.70.205333}{\bibinfo {journal} {{P}hys. ~{R}ev. ~{B}}}\
  }%
  \textbf{\bibinfo {volume} {70}},\ \bibinfo {pages} {205333} (\bibinfo {year}
  {2004})%
  \bibAnnoteFile{NoStop}{DigammaBulka}%
\bibitem{Johnson1992}%
  \BibitemOpen
  \bibfield{author}{%
  \bibinfo {author} {\bibfnamefont{A.~T.}\ \bibnamefont{Johnson}}, \bibinfo
  {author} {\bibfnamefont{L.~P.}\ \bibnamefont{Kouwenhoven}}, \bibinfo {author}
  {\bibfnamefont{W.}~\bibnamefont{de~Jong}}, \bibinfo {author}
  {\bibfnamefont{N.~C.}\ \bibnamefont{van~der Vaart}}, \bibinfo {author}
  {\bibfnamefont{C.~J. P.~M.}\ \bibnamefont{Harmans}},\ and\ \bibinfo {author}
  {\bibfnamefont{C.~T.}\ \bibnamefont{Foxon}},\ }%
  \bibfield{journal}{%
  \Doi{10.1103/PhysRevLett.69.1592}{\bibinfo {journal}
  {{P}hys.~{R}ev.~{L}ett.}}\ }%
  \textbf{\bibinfo {volume} {69}},\ \bibinfo {pages} {1592} (\bibinfo {year}
  {1992})%
  \bibAnnoteFile{NoStop}{Johnson1992}%
\bibitem{Foxman1993}%
  \BibitemOpen
  \bibfield{author}{%
  \bibinfo {author} {\bibfnamefont{E.~B.}\ \bibnamefont{Foxman}}, \bibinfo
  {author} {\bibfnamefont{P.~L.}\ \bibnamefont{McEuen}}, \bibinfo {author}
  {\bibfnamefont{U.}~\bibnamefont{Meirav}}, \bibinfo {author}
  {\bibfnamefont{N.}~\bibnamefont{S.Wingreen}}, \bibinfo {author}
  {\bibfnamefont{Y.}~\bibnamefont{Meir}}, \bibinfo {author}
  {\bibfnamefont{P.~A.}\ \bibnamefont{Belk}}, \bibinfo {author}
  {\bibfnamefont{N.~R.}\ \bibnamefont{Belk}},\ and\ \bibinfo {author}
  {\bibfnamefont{M.~A.}\ \bibnamefont{Kastner}},\ }%
  \bibfield{journal}{%
  \Doi{10.1103/PhysRevB.47.10020}{\bibinfo {journal} {{P}hys.~{R}ev.~B}}\ }%
  \textbf{\bibinfo {volume} {47}},\ \bibinfo {pages} {10020} (\bibinfo {year}
  {1993})%
  \bibAnnoteFile{NoStop}{Foxman1993}%
\bibitem{Reddy2007}%
  \BibitemOpen
  \bibfield{author}{%
  \bibinfo {author} {\bibfnamefont{P.}~\bibnamefont{Reddy}}, \bibinfo {author}
  {\bibfnamefont{S.-Y.}\ \bibnamefont{Jang}}, \bibinfo {author}
  {\bibfnamefont{R.~A.}\ \bibnamefont{Segalman}},\ and\ \bibinfo {author}
  {\bibfnamefont{A.}~\bibnamefont{Majumdar}},\ }%
  \bibfield{journal}{%
  \Doi{10.1126/science.1137149}{\bibinfo {journal} {Science}}\ }%
  \textbf{\bibinfo {volume} {315}},\ \bibinfo {pages} {1568} (\bibinfo {year}
  {2007})%
  \bibAnnoteFile{NoStop}{Reddy2007}%
\bibitem{CNTSET}%
  \BibitemOpen
  \bibfield{author}{%
  \bibinfo {author} {\bibfnamefont{H.~W.~C.}\ \bibnamefont{Postma}}, \bibinfo
  {author} {\bibfnamefont{T.}~\bibnamefont{Teepen}}, \bibinfo {author}
  {\bibfnamefont{Z.}~\bibnamefont{Yao}}, \bibinfo {author}
  {\bibfnamefont{M.}~\bibnamefont{Grifoni}},\ and\ \bibinfo {author}
  {\bibfnamefont{C.}~\bibnamefont{Dekker}},\ }%
  \bibfield{journal}{%
  \Doi{10.1126/science.1061797}{\bibinfo {journal} {Science}}\ }%
  \textbf{\bibinfo {volume} {293}},\ \bibinfo {pages} {76} (\bibinfo {year}
  {2001})%
  \bibAnnoteFile{NoStop}{CNTSET}%
\bibitem{Linke2012}%
  \BibitemOpen
  \bibfield{author}{%
  \bibinfo {author} {\bibfnamefont{S.~F.}\ \bibnamefont{Svensson}}, \bibinfo
  {author} {\bibfnamefont{A.~I.}\ \bibnamefont{Persson}}, \bibinfo {author}
  {\bibfnamefont{E.~A.}\ \bibnamefont{Hoffmann}}, \bibinfo {author}
  {\bibfnamefont{N.}~\bibnamefont{Nakpathomkun}}, \bibinfo {author}
  {\bibfnamefont{H.~A.}\ \bibnamefont{Nilsson}}, \bibinfo {author}
  {\bibfnamefont{H.~Q.}\ \bibnamefont{Xu}}, \bibinfo {author}
  {\bibfnamefont{L.}~\bibnamefont{Samuelson}},\ and\ \bibinfo {author}
  {\bibfnamefont{H.}~\bibnamefont{Linke}},\ }%
  \bibfield{journal}{%
  \Doi{10.1088/1367-2630/14/3/033041}{\bibinfo {journal} {{N}ew. ~{J}.
  ~{P}hys}}\ }%
  \textbf{\bibinfo {volume} {14}},\ \bibinfo {pages} {033041} (\bibinfo {year}
  {2012})%
  \bibAnnoteFile{NoStop}{Linke2012}%
\bibitem{Chen2010}%
  \BibitemOpen
  \bibfield{author}{%
  \bibinfo {author} {\bibfnamefont{J.}~\bibnamefont{Chen}}, \bibinfo {author}
  {\bibfnamefont{T.}~\bibnamefont{Markussen}},\ and\ \bibinfo {author}
  {\bibfnamefont{K.~S.}\ \bibnamefont{Thygesen}},\ }%
  \bibfield{journal}{%
  \Doi{10.1103/PhysRevB.82.121412}{\bibinfo {journal} {{P}hys.~{R}ev.~{B}}}\ }%
  \textbf{\bibinfo {volume} {82}},\ \bibinfo {pages} {121412(R)} (\bibinfo
  {year} {2010})%
  \bibAnnoteFile{NoStop}{Chen2010}%
\bibitem{Stafford1996}%
  \BibitemOpen
  \bibfield{author}{%
  \bibinfo {author} {\bibfnamefont{C.~A.}\ \bibnamefont{Stafford}},\ }%
  \bibfield{journal}{%
  \Doi{10.1103/PhysRevLett.77.2770}{\bibinfo {journal}
  {{P}hys.~{R}ev.~{L}ett.}}\ }%
  \textbf{\bibinfo {volume} {77}},\ \bibinfo {pages} {2770} (\bibinfo {year}
  {1996})%
  \bibAnnoteFile{NoStop}{Stafford1996}%
\bibitem{Christen1994}%
  \BibitemOpen
  \bibfield{author}{%
  \bibinfo {author} {\bibfnamefont{T.}~\bibnamefont{Christen}}\ and\ \bibinfo
  {author} {\bibfnamefont{M.}~\bibnamefont{B$\ddot{\mathrm{u}}$ttiker}},\ }%
  \bibfield{journal}{%
  \bibinfo {journal} {{E}urophys.~{L}ett.}\ }%
  \textbf{\bibinfo {volume} {35}},\ \bibinfo {pages} {523} (\bibinfo {year}
  {1996})%
  \bibAnnoteFile{NoStop}{Christen1994}%
\bibitem{Logan2001}%
  \BibitemOpen
  \bibfield{author}{%
  \bibinfo {author} {\bibfnamefont{N.~L.}\ \bibnamefont{Dickens}}\ and\
  \bibinfo {author} {\bibfnamefont{D.~E.}\ \bibnamefont{Logan}},\ }%
  \bibfield{journal}{%
  \Doi{10.1088/0953-8984/13/20/311}{\bibinfo {journal} {{J}. ~{P}hys:.
  ~{C}ondens. ~{M}atter}}\ }%
  \textbf{\bibinfo {volume} {13}},\ \bibinfo {pages} {4505} (\bibinfo {year}
  {2001})%
  \bibAnnoteFile{NoStop}{Logan2001}%
\bibitem{Stegun}%
  \BibitemOpen
  \bibfield{author}{%
  \bibinfo {author} {\bibfnamefont{M.}~\bibnamefont{Abramowitz}}\ and\ \bibinfo
  {author} {\bibfnamefont{I.~A.}\ \bibnamefont{Stegun}},\ }%
  \emph{\bibinfo {title} {Handbook of Mathematical Functions with Formulas,
  Graphs, and Mathematical Tables}},\ \bibinfo {edition} {ninth dover printing,
  tenth gpo printing}\ ed.\ (\bibinfo {publisher} {Dover},\ \bibinfo {address}
  {New York},\ \bibinfo {year} {1964})%
  \bibAnnoteFile{NoStop}{Stegun}%
\bibitem{Subroto2008}%
  \BibitemOpen
  \bibfield{author}{%
  \bibinfo {author} {\bibfnamefont{P.}~\bibnamefont{Murphy}}, \bibinfo {author}
  {\bibfnamefont{S.}~\bibnamefont{Mukerjee}}, ,\ and\ \bibinfo {author}
  {\bibfnamefont{J.}~\bibnamefont{Moore}},\ }%
  \bibfield{journal}{%
  \Doi{10.1103/PhysRevB.78.161406}{\bibinfo {journal} {{P}hys.~{R}ev.~{B}}}\ }%
  \textbf{\bibinfo {volume} {78}},\ \bibinfo {pages} {161406(R)} (\bibinfo
  {year} {2008})%
  \bibAnnoteFile{NoStop}{Subroto2008}%
\bibitem{Rejec2012}%
  \BibitemOpen
  \bibfield{author}{%
  \bibinfo {author} {\bibfnamefont{T.}~\bibnamefont{Rejec}}, \bibinfo {author}
  {\bibfnamefont{R.}~\bibnamefont{Zitko}}, \bibinfo {author}
  {\bibfnamefont{J.}~\bibnamefont{Mravlje}},\ and\ \bibinfo {author}
  {\bibfnamefont{A.}~\bibnamefont{Ramsak}},\ }%
  \bibfield{journal}{%
  \Doi{10.1103/PhysRevB.85.085117}{\bibinfo {journal} {{P}hys.~{R}ev.~{B}}}\ }%
  \textbf{\bibinfo {volume} {85}},\ \bibinfo {pages} {085117} (\bibinfo {year}
  {2012})%
  \bibAnnoteFile{NoStop}{Rejec2012}%
\bibitem{identities_SI}%
  \BibitemOpen
  \bibinfo {note} {See Supplemental Material at [URL will be inserted by
  publisher] for the identites provided in Section IV}%
  \bibAnnoteFile{NoStop}{identities_SI}%
\bibitem{KopninGalperin2009}%
  \BibitemOpen
  \bibfield{author}{%
  \bibinfo {author} {\bibfnamefont{N.~B.}\ \bibnamefont{Kopnin}}, \bibinfo
  {author} {\bibfnamefont{Y.~M.}\ \bibnamefont{Galperin}},\ and\ \bibinfo
  {author} {\bibfnamefont{V.~M.}\ \bibnamefont{Vinokur}},\ }%
  \bibfield{journal}{%
  \Doi{10.1103/PhysRevB.79.035319}{\bibinfo {journal} {{P}hys.~{R}ev.~{B}}}\ }%
  \textbf{\bibinfo {volume} {79}},\ \bibinfo {pages} {035319} (\bibinfo {year}
  {2009})%
  \bibAnnoteFile{NoStop}{KopninGalperin2009}%
\bibitem{Beenakker1991}%
  \BibitemOpen
  \bibfield{author}{%
  \bibinfo {author} {\bibfnamefont{C.~W.~J.}\ \bibnamefont{Beenakker}},\ }%
  \bibfield{journal}{%
  \Doi{10.1103/PhysRevB.44.1646}{\bibinfo {journal} {{P}hys. ~{R}ev. ~{B}}}\ }%
  \textbf{\bibinfo {volume} {44}},\ \bibinfo {pages} {1646} (\bibinfo {year}
  {1991})%
  \bibAnnoteFile{NoStop}{Beenakker1991}%
\bibitem{Linke2011}%
  \BibitemOpen
  \bibfield{author}{%
  \bibinfo {author} {\bibfnamefont{P.}~\bibnamefont{Mani}}, \bibinfo {author}
  {\bibfnamefont{N.}~\bibnamefont{Nakpathomkun}}, \bibinfo {author}
  {\bibfnamefont{E.~A.}\ \bibnamefont{Hoffmann}},\ and\ \bibinfo {author}
  {\bibfnamefont{H.}~\bibnamefont{Linke}},\ }%
  \bibfield{journal}{%
  \Doi{10.1021/nl202258f}{\bibinfo {journal} {{N}ano~{L}ett.}}\ }%
  \textbf{\bibinfo {volume} {11}},\ \bibinfo {pages} {4679} (\bibinfo {year}
  {2011})%
  \bibAnnoteFile{NoStop}{Linke2011}%
\bibitem{NR_SI}%
  \BibitemOpen
  \bibinfo {note} {See Supplemental Material at [URL will be inserted by
  publisher] for detailed derivations and expressions and additional figures
  relevant for the narrow resonance regime in Section II}%
  \bibAnnoteFile{NoStop}{NR_SI}%
\bibitem{MATLAB}%
  \BibitemOpen
  \bibfield{author}{%
  \bibinfo {author} {\bibnamefont{MATLAB}},\ }%
  \emph{\bibinfo {title} {version 7.10.0 (R2010a)}}\ (\bibinfo {publisher} {The
  MathWorks Inc.},\ \bibinfo {address} {Natick, Massachusetts},\ \bibinfo
  {year} {2010})%
  \bibAnnoteFile{NoStop}{MATLAB}%
\bibitem{NL_parameters}%
  \BibitemOpen
  \bibinfo {note} {These parameters are justified within our theory and are
  experimentally feasible within the specified approximations, for e.g see
  \cite{Linke2012}.}%
  \bibAnnoteFile{Stop}{NL_parameters}%
\bibitem{Ioan2012}%
  \BibitemOpen
  \bibfield{author}{%
  \bibinfo {author} {\bibfnamefont{I.}~\bibnamefont{B$\hat{a}$ldea}},\ }%
  \bibfield{journal}{%
  \Doi{10.1103/PhysRevB.85.035442}{\bibinfo {journal} {{P}hys.~{R}ev.~{B}}}\ }%
  \textbf{\bibinfo {volume} {85}},\ \bibinfo {pages} {035442} (\bibinfo {year}
  {2012})%
  \bibAnnoteFile{NoStop}{Ioan2012}%
\bibitem{BR_SI}%
  \BibitemOpen
  \bibinfo {note} {See Supplemental Material at [URL will be inserted by
  publisher] for detailed derivations and expressions and additional figures
  relevant for the broad resonance regime in Section III}%
  \bibAnnoteFile{NoStop}{BR_SI}%
\bibitem{TVS1Yu}%
  \BibitemOpen
  \bibfield{author}{%
  \bibinfo {author} {\bibfnamefont{L.~H.}\ \bibnamefont{Yu}}, \bibinfo {author}
  {\bibfnamefont{N.}~\bibnamefont{Gergel-Hackett}}, \bibinfo {author}
  {\bibfnamefont{C.~D.}\ \bibnamefont{Zangmeister}}, \bibinfo {author}
  {\bibfnamefont{C.~A.}\ \bibnamefont{Hacker}}, \bibinfo {author}
  {\bibfnamefont{C.~A.}\ \bibnamefont{Richter}},\ and\ \bibinfo {author}
  {\bibfnamefont{J.~G.}\ \bibnamefont{Kushmerick}},\ }%
  \bibfield{journal}{%
  \bibinfo {journal} {{J}.~ {P}hys.~ {C}ondens.~{M}atter}\ }%
  \textbf{\bibinfo {volume} {20}},\ \bibinfo {pages} {374114} (\bibinfo {year}
  {2008})%
  \bibAnnoteFile{NoStop}{TVS1Yu}%
\bibitem{TVS2Liu}%
  \BibitemOpen
  \bibfield{author}{%
  \bibinfo {author} {\bibfnamefont{K.}~\bibnamefont{Liu}}, \bibinfo {author}
  {\bibfnamefont{X.}~\bibnamefont{Wang}},\ and\ \bibinfo {author}
  {\bibfnamefont{F.}~\bibnamefont{Wang}},\ }%
  \bibfield{journal}{%
  \bibinfo {journal} {{A}CS~{N}ano}\ }%
  \textbf{\bibinfo {volume} {2}},\ \bibinfo {pages} {2315} (\bibinfo {year}
  {2008})%
  \bibAnnoteFile{NoStop}{TVS2Liu}%
\bibitem{Beebe2006}%
  \BibitemOpen
  \bibfield{author}{%
  \bibinfo {author} {\bibfnamefont{J.~M.}\ \bibnamefont{Beebe}}, \bibinfo
  {author} {\bibfnamefont{B.}~\bibnamefont{Kim}}, \bibinfo {author}
  {\bibfnamefont{J.~W.}\ \bibnamefont{Gadzuk}}, \bibinfo {author}
  {\bibfnamefont{C.~D.}\ \bibnamefont{Frisbie}},\ and\ \bibinfo {author}
  {\bibfnamefont{J.~G.}\ \bibnamefont{Kushmerick}},\ }%
  \bibfield{journal}{%
  \Doi{10.1103/PhysRevLett.97.026801}{\bibinfo {journal}
  {{P}hys.~{R}ev.~{L}ett.}}\ }%
  \textbf{\bibinfo {volume} {97}},\ \bibinfo {pages} {026801} (\bibinfo {year}
  {2006})%
  \bibAnnoteFile{NoStop}{Beebe2006}%
\bibitem{Beebe2008}%
  \BibitemOpen
  \bibfield{author}{%
  \bibinfo {author} {\bibfnamefont{J.~M.}\ \bibnamefont{Beebe}}, \bibinfo
  {author} {\bibfnamefont{B.}~\bibnamefont{Kim}}, \bibinfo {author}
  {\bibfnamefont{C.~D.}\ \bibnamefont{Frisbie}},\ and\ \bibinfo {author}
  {\bibfnamefont{J.~G.}\ \bibnamefont{Kushmerick}},\ }%
  \bibfield{journal}{%
  \Doi{10.1021/nn700424u}{\bibinfo {journal} {{A}CS~{N}ano}}\ }%
  \textbf{\bibinfo {volume} {2}},\ \bibinfo {pages} {827} (\bibinfo {year}
  {2008})%
  \bibAnnoteFile{NoStop}{Beebe2008}%
\bibitem{Tan2010}%
  \BibitemOpen
  \bibfield{author}{%
  \bibinfo {author} {\bibfnamefont{A.}~\bibnamefont{Tan}}, \bibinfo {author}
  {\bibfnamefont{S.}~\bibnamefont{Sadat}},\ and\ \bibinfo {author}
  {\bibfnamefont{P.}~\bibnamefont{Reddy}},\ }%
  \bibfield{journal}{%
  \Doi{10.1063/1.3291521}{\bibinfo {journal} {{A}ppl.~{Phys}.~{L}ett.}}\ }%
  \textbf{\bibinfo {volume} {96}},\ \bibinfo {pages} {013110} (\bibinfo {year}
  {2010})%
  \bibAnnoteFile{NoStop}{Tan2010}%
\bibitem{Hybertsen2008}%
  \BibitemOpen
  \bibfield{author}{%
  \bibinfo {author} {\bibfnamefont{M.~S.}\ \bibnamefont{Hybertsen}}, \bibinfo
  {author} {\bibfnamefont{L.}~\bibnamefont{Venkataraman}}, \bibinfo {author}
  {\bibfnamefont{J.~E.}\ \bibnamefont{Klare}}, \bibinfo {author}
  {\bibfnamefont{A.}~\bibnamefont{CWhalley}}, \bibinfo {author}
  {\bibfnamefont{M.~L.}\ \bibnamefont{Steigerwald}},\ and\ \bibinfo {author}
  {\bibfnamefont{C.}~\bibnamefont{Nuckolls}},\ }%
  \bibfield{journal}{%
  \Doi{10.1088/0953-8984/20/37/374115}{\bibinfo {journal} {{J}. ~{P}hys:
  ~{C}ondens. ~{M}atter}}\ }%
  \textbf{\bibinfo {volume} {20}},\ \bibinfo {pages} {374115} (\bibinfo {year}
  {2008})%
  \bibAnnoteFile{NoStop}{Hybertsen2008}%
\bibitem{Hybertsen2007}%
  \BibitemOpen
  \bibfield{author}{%
  \bibinfo {author} {\bibfnamefont{S.~Y.}\ \bibnamefont{Quek}}, \bibinfo
  {author} {\bibfnamefont{L.}~\bibnamefont{Venkataraman}}, \bibinfo {author}
  {\bibfnamefont{H.~J.}\ \bibnamefont{Choi}}, \bibinfo {author}
  {\bibfnamefont{S.~G.}\ \bibnamefont{Louie}}, \bibinfo {author}
  {\bibfnamefont{M.~S.}\ \bibnamefont{Hybertsen}},\ and\ \bibinfo {author}
  {\bibfnamefont{J.~B.}\ \bibnamefont{Neaton}},\ }%
  \bibfield{journal}{%
  \Doi{10.1021/nl072058i}{\bibinfo {journal} {{N}ano~{L}ett.}}\ }%
  \textbf{\bibinfo {volume} {7}},\ \bibinfo {pages} {3477} (\bibinfo {year}
  {2007})%
  \bibAnnoteFile{NoStop}{Hybertsen2007}%
\bibitem{FW}%
  \BibitemOpen
  \bibinfo {note} {See Supplemental Material at [URL will be inserted by
  publisher] for the derivation.}%
  \bibAnnoteFile{Stop}{FW}%
\bibitem{Sanvito2012}%
  \BibitemOpen
  \bibfield{author}{%
  \bibinfo {author} {\bibfnamefont{J.-X.}\ \bibnamefont{Yu}}, \bibinfo {author}
  {\bibfnamefont{X.-R.}\ \bibnamefont{Chen}}, \bibinfo {author}
  {\bibfnamefont{S.}~\bibnamefont{Sanvito}},\ and\ \bibinfo {author}
  {\bibfnamefont{Y.}~\bibnamefont{Cheng}},\ }%
  \bibfield{journal}{%
  \Doi{10.1063/1.3665614}{\bibinfo {journal} {{A}ppl.~{Phys}.~{L}ett.}}\ }%
  \textbf{\bibinfo {volume} {100}},\ \bibinfo {pages} {013113} (\bibinfo {year}
  {2012})%
  \bibAnnoteFile{NoStop}{Sanvito2012}%
\bibitem{benzene}%
  \BibitemOpen
  \bibfield{author}{%
  \bibinfo {author} {\bibfnamefont{Y.}~\bibnamefont{Li}}, \bibinfo {author}
  {\bibfnamefont{P.}~\bibnamefont{Wei}}, \bibinfo {author}
  {\bibfnamefont{M.}~\bibnamefont{Bai}}, \bibinfo {author}
  {\bibfnamefont{Z.}~\bibnamefont{Shen}}, \bibinfo {author}
  {\bibfnamefont{S.}~\bibnamefont{Sanvito}},\ and\ \bibinfo {author}
  {\bibfnamefont{S.}~\bibnamefont{Hou}},\ }%
  \bibfield{journal}{%
  \bibinfo {journal} {{C}hem.~{P}hys.}\ }%
  \textbf{\bibinfo {volume} {397}},\ \bibinfo {pages} {82} (\bibinfo {year}
  {2012})%
  \bibAnnoteFile{NoStop}{benzene}%
\bibitem{Kotliar2010}%
  \BibitemOpen
  \bibfield{author}{%
  \bibinfo {author} {\bibfnamefont{D.}~\bibnamefont{Jacob}}, \bibinfo {author}
  {\bibfnamefont{K.}~\bibnamefont{Haule}},\ and\ \bibinfo {author}
  {\bibfnamefont{G.}~\bibnamefont{Kotliar}},\ }%
  \bibfield{journal}{%
  \Doi{10.1103/PhysRevB.82.195115}{\bibinfo {journal} {{P}hys.~{R}ev.~{B}}}\ }%
  \textbf{\bibinfo {volume} {82}},\ \bibinfo {pages} {195115} (\bibinfo {year}
  {2010})%
  \bibAnnoteFile{NoStop}{Kotliar2010}%
\bibitem{Markussennanolett}%
  \BibitemOpen
  \bibfield{author}{%
  \bibinfo {author} {\bibfnamefont{T.}~\bibnamefont{Markussen}}, \bibinfo
  {author} {\bibfnamefont{R.}~\bibnamefont{Stadler}},\ and\ \bibinfo {author}
  {\bibfnamefont{K.~S.}\ \bibnamefont{Thygesen}},\ }%
  \bibfield{journal}{%
  \Doi{10.1021/nl101688a}{\bibinfo {journal} {{N}ano~{L}ett.}}\ }%
  \textbf{\bibinfo {volume} {10}},\ \bibinfo {pages} {4260} (\bibinfo {year}
  {2010})%
  \bibAnnoteFile{NoStop}{Markussennanolett}%
\bibitem{Stafford2009}%
  \BibitemOpen
  \bibfield{author}{%
  \bibinfo {author} {\bibfnamefont{J.~P.}\ \bibnamefont{Bergfield}}\ and\
  \bibinfo {author} {\bibfnamefont{C.~A.}\ \bibnamefont{Stafford}},\ }%
  \bibfield{journal}{%
  \Doi{10.1103/PhysRevB.79.245125}{\bibinfo {journal} {{P}hys.~{R}ev.~{B}}}\ }%
  \textbf{\bibinfo {volume} {79}},\ \bibinfo {pages} {245125} (\bibinfo {year}
  {2009})%
  \bibAnnoteFile{NoStop}{Stafford2009}%
\bibitem{DiVentra}%
  \BibitemOpen
  \bibfield{author}{%
  \bibinfo {author} {\bibfnamefont{M.}~\bibnamefont{Brandbyge}}, \bibinfo
  {author} {\bibfnamefont{J.-L.}\ \bibnamefont{Mozos}}, \bibinfo {author}
  {\bibfnamefont{P.}~\bibnamefont{Ordejon}}, \bibinfo {author}
  {\bibfnamefont{J.}~\bibnamefont{Taylor}},\ and\ \bibinfo {author}
  {\bibfnamefont{K.}~\bibnamefont{Stokbro}},\ }%
  \bibfield{journal}{%
  \Doi{10.1103/PhysRevB.65.165401}{\bibinfo {journal} {{P}hys. ~{R}ev. ~{B}}}\
  }%
  \textbf{\bibinfo {volume} {65}},\ \bibinfo {pages} {165401} (\bibinfo {year}
  {2002})%
  \bibAnnoteFile{NoStop}{DiVentra}%
\end{thebibliography}%

\end{document}